\documentclass{iau} 
\usepackage{graphicx}
\usepackage{natbib,twoopt}
\usepackage[breaklinks=true]{hyperref}
\bibpunct{(}{)}{;}{a}{}{,} 

\title[Multiple populations in globular clusters: spectroscopy] %% give here short title %%
{Spectroscopic studies of stellar populations in globular clusters and field stars: implications for globular cluster and Milky Way halo formation}

\author[R. Gratton]   %% give here short author list %%
{Raffaele Gratton
        }

\affiliation{
INAF-Osservatorio Astronomico di Padova,
}

\pubyear{2019}
\volume{351}  
\jname{Star Clusters: From the Milky Way to the Early Universe}
\editors{A. Bragaglia, M.B. Davies, A. Sills \& E. Vesperini, eds.}
\begin{document}

\maketitle
%. CONTINUE EDITING FROM HERE

\begin{abstract}
We review spectroscopic results concerning multiple stellar populations in globular clusters. The cluster initial mass is the most important parameter determining the fraction of second generation stars. The threshold for the onset of the multiple population phenomenon is 1-$3\times 10^5$~M$_\odot$. Nucleosynthesis is influenced by metallicity: Na/O and Mg/Al anti-correlations are more extended in metal-poor than in metal-rich clusters. Massive clusters are more complex systems than the smaller ones, with several populations characterized by different chemical compositions. The high Li abundance observed in the intermediate second generation stars strongly favours intermediate mass AGB stars as polluters for this class of stars; however, it is well possible that the polluters of extreme second generation stars, that often do not have measurable Li, may be fast rotating massive stars or super-massive stars. The mass budget factor should be a function of the cluster mass, and needs to be large only in massive clusters.

\end{abstract}

\firstsection % if your document starts with a section,
\section{Introduction}

A basic result of the last twenty years is that globular clusters host multiple stellar populations \citep{Gratton2001}, differing according to their chemical composition (for reviews, see \citealt{Gratton2004, Gratton2012, Bastian2018}). This is very likely connected to how they formed, but the exact relation is currently unknown. The different populations may be studied through spectroscopy and precision photometry. As discussed recently by \citet{Marino2019}, there is substantial agreement between results obtained through the two methods. Photometry allows a much wider statistics and possibly a cleaner distinction between the different populations present in clusters, while spectroscopy gives more insight into the nucleosynthesis mechanisms, providing information on many different elements. Combined, they are very powerful tools to discuss the properties of the multiple populations in globular clusters. Here we review some recent findings, with emphasis on those that were obtained through spectroscopy; the accompanying review by Milone rather emphasizes results obtained through photometry. 

We will only touch a few points here; a more extensive discussion may be found in a paper we are preparing for Astronomy $\&$ Astrophysics Review (Gratton et al.  2019, submitted). In particular, we will briefly present the correlations between photometric indices and spectroscopy (see also the contribution by Marino at this meeting); discuss the relations with cluster masses, metallicity, and location in the Milky Way; give some update about Lithium; and briefly comment on the mass budget issue. These are only a fraction of the most relevant topics; we refer to the above mentioned review for a more extensive discussion.

\section{Anti-correlations}

Chemical inhomogeneities in globular clusters show up as correlations and anti-correlations between the abundances of different elements: C is anti-correlated with N, Na with O, Mg with Al (and in some case with Si), and in a very few cases, Ca with K. While the two first anti-correlations are ubiquitous in globular clusters, the others are present only in a fraction of them. They are all present only in a couple of cases (NGC~2808 and NGC~2419).

These anti-correlations are related to H-burning at high temperature \citep{Denisenkov1989, Langer1993, Prantzos2017}. The various anti-correlations require different burning temperatures $T$: C-N: $T\sim$10 MK; Na/O: $T\sim$40 MK; Mg/Al: $T\sim$70 MK; Al-Si: $T\sim$80 MK; K-Ca: $T\sim$180 MK. However, good polluter candidates not only need that in some region H is burnt at such high temperatures, but also that the products of the burning are brought at the surface of the star by some mixing and then lost to the interstellar medium by a suitable mass loss mechanism. Mixing and mass loss are then basic aspects to be considered. We expect that all these features can be reached in different objects, that can then be considered as possible polluters. The most popular in the present context are fast rotating massive stars \citep{Decressin2007}, intermediate mass AGB stars \citep{Ventura2001}, massive binaries \citep{deMink2009}, and supermassive stars \citep{Denissenkov2014, Gieles2018}.

In addition, the anti-correlations likely require dilution with material with pristine chemical composition (see e.g. \citealt{Prantzos2006}) to reproduce what is observed. This is most evident for the case of the Na/O anti-correlation. This diluting material may be leftover from the formation phase (see e.g. \citealt{Dercole2008, Bekki2007}) or may come from less evolved single \citep{Gratton2011b} or binary stars \citep{Vanbeveren2012}. 
As discussed by \citet{Renzini2015} and \citet{Bastian2018}, scenarios appropriate to all these different candidate polluters meet serious difficulties in explaining the whole set of observations when they are considered individually. While we have no room here to enter into the details of this discussion, we should be open to the possibility that the explanation is complex, with different mechanisms possibly being active and explaining different classes of second generation objects.

A main step forward in our description of the multiple populations in globular clusters was made with the introduction of the so-called chromosome diagram \citep{Milone2017}. This was obtained exploiting UV photometry provided by the Hubble Space Telescope. In the chromosome diagram, each star in the cluster is represented by a point. In most cases, for simplicity, only stars on the lower red giant branch are considered, but as shown e.g. by \citet{Milone2012a}, it is possible to separate stars of the different populations throughout most evolutionary phases using similar diagrams. The x-axis of the diagram is the pseudo-colour $\Delta_{F275W,F814W}$, that is essentially determined by the He content; the y-axis is the pseudo-colour difference $\Delta_{C~{\rm F275W,F336W,F438W}}$, that is essentially determined by the N content. The power of this diagram is that stars in a globular cluster typically divide into two well separated groups: a group at $\Delta_{C~{\rm F275W,F336W,F438W}} \sim 0$, that is, N-poor stars, usually called first generation (FG) stars; and a group with $\Delta_{C~{\rm F275W,F336W,F438W}} > 0.1$, that is, N-rich stars, usually called second generation (SG) stars. In most clusters, only two groups are obvious in the chromosome diagram. However, in a few clusters there are additional groups of stars, typically differing in the value of $\Delta_{\rm F275W,F814W}$ at a given $\Delta_{C~{\rm F275W,F336W,F438W}}$ value. Clusters with chromosome diagram with only two groups of stars are called of Type I; those with more groups are called of type II. Type II clusters have a more complex pattern of chemical inhomogeneities and presumably a more complex history. 

\begin{figure}[ht]
%\vspace*{-0.6 cm}
\begin{center}
\includegraphics[width=3.4in]{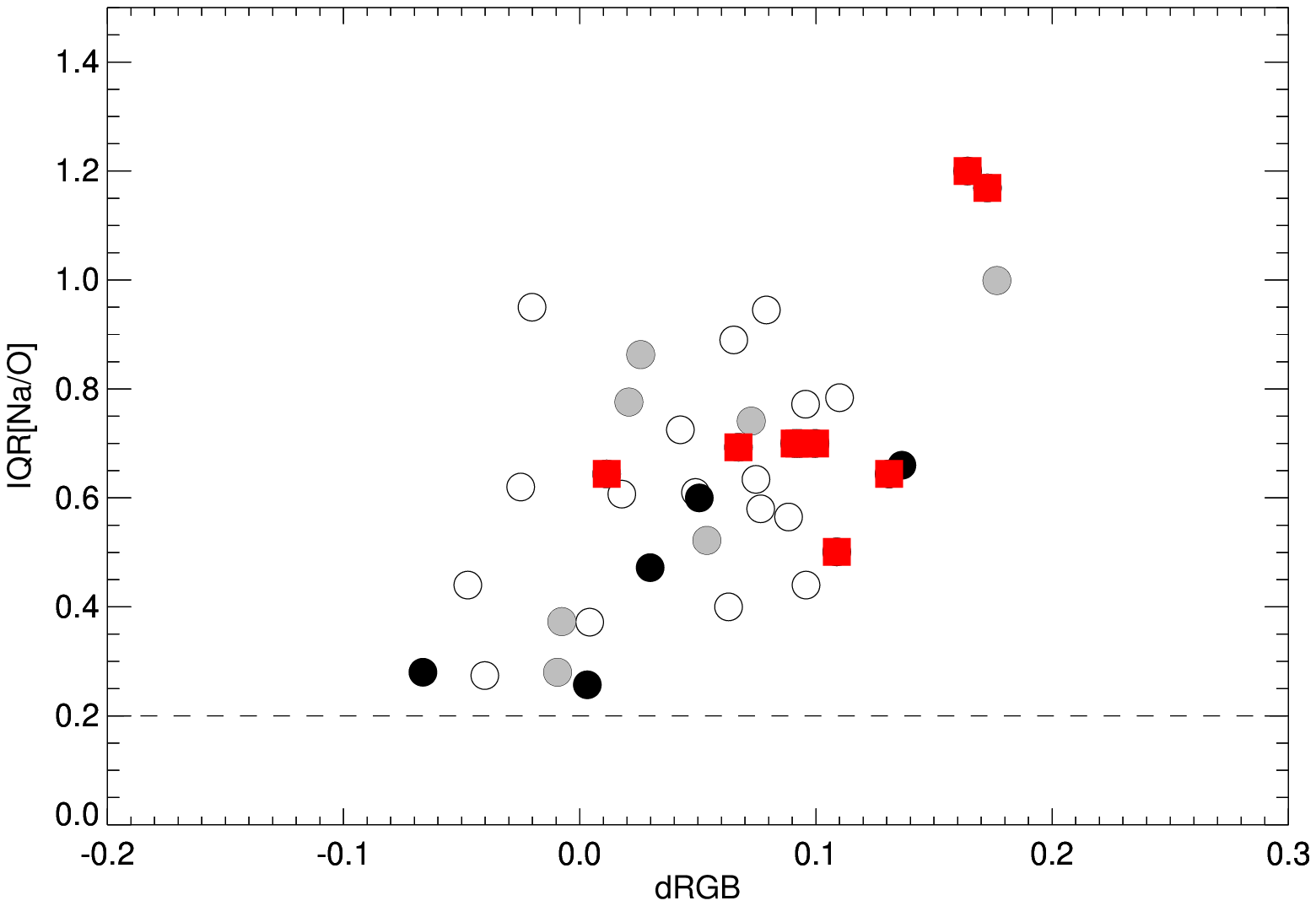}
\includegraphics[width=3.4in]{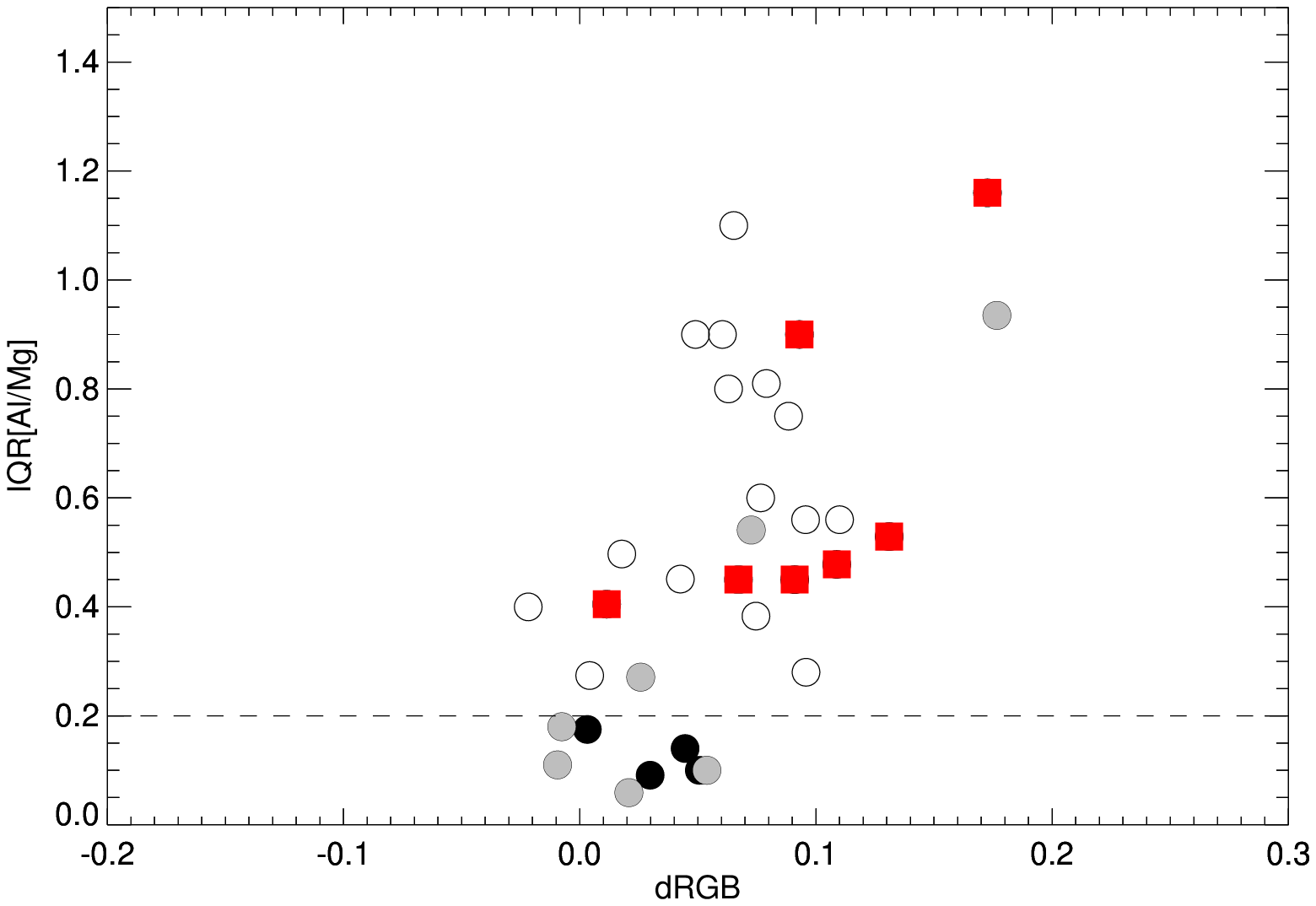}
% \vspace*{-1.0 cm}
\caption{Upper panel: Correlation between the photometric index dRGB=$\Delta(275W,F336W,F438W)$-ref from \citet{Milone2017} -  measuring the spread in N abundance within a cluster - and the interquartile of the [Na/O] values (IQR[Na/O]) obtained from spectroscopy (see Gratton et al., 2019, submitted). Each point is the value obtained for an individual cluster. Red squares are type II clusters; open circles are type I clusters with [Fe/H]$<-1.5$, grey filled circles are type I clusters with -1.5$<$[Fe/H]$<$-1.0, and open circles are type I clusters with [Fe/H]$>-1.0$. IQR values below the dashed line are compatible with no real spread in the abundances within a cluster. Lower panel: the same, but for the interquartile of the [Mg/Al] ratio IQR[Mg/Al].}
\label{f:fig1}
\end{center}
\end{figure}

\section{Multiple populations and main cluster parameters}

Confirming and extending previous results (see e.g. \citealp{Carretta2010}), recently \citet{Marino2019} showed that there is a close star-to-star correlation between the FG/SG classification based on the chromosome diagram and on the Na/O and Mg/Al anti-correlations. Here, we will show some examples of the correlations existing between the spread in the various indices and chemical abundances among different globular clusters. For this purpose, we will use the index dRGB=$\Delta(275W,F336W,F438W)$\ from \citet{Milone2017}, that is measuring the variation in N abundance, and the interquartiles IQR(Na/O) and IQR(Mg/Al) from our FLAMES survey of 25 globular clusters and a collection of other literature data (for details, see Gratton et al. 2019, submitted). The correlations between these different quantities are shown in Figure~\ref{f:fig1}. We used different symbols for type I and II clusters, and for type I clusters with different metallicity.

This figure shows that there is a good correlation between the spread obtained for N from photometry and those obtained along the Na/O and Mg/Al anti-correlations. However, there is a systematic trend for metal-rich clusters to have smaller spread in Na/O and moreover in Mg/Al than obtained for N abundances. This confirms early findings by \citet{Carretta2009b} and others (see e.g. \citealt{Nataf2019}) that the nucleosynthesis involved in the multiple population phenomenon is sensitive to the metallicity.

\begin{figure}[ht]
%\vspace*{-0.6 cm}
\begin{center}
\includegraphics[width=3.4in]{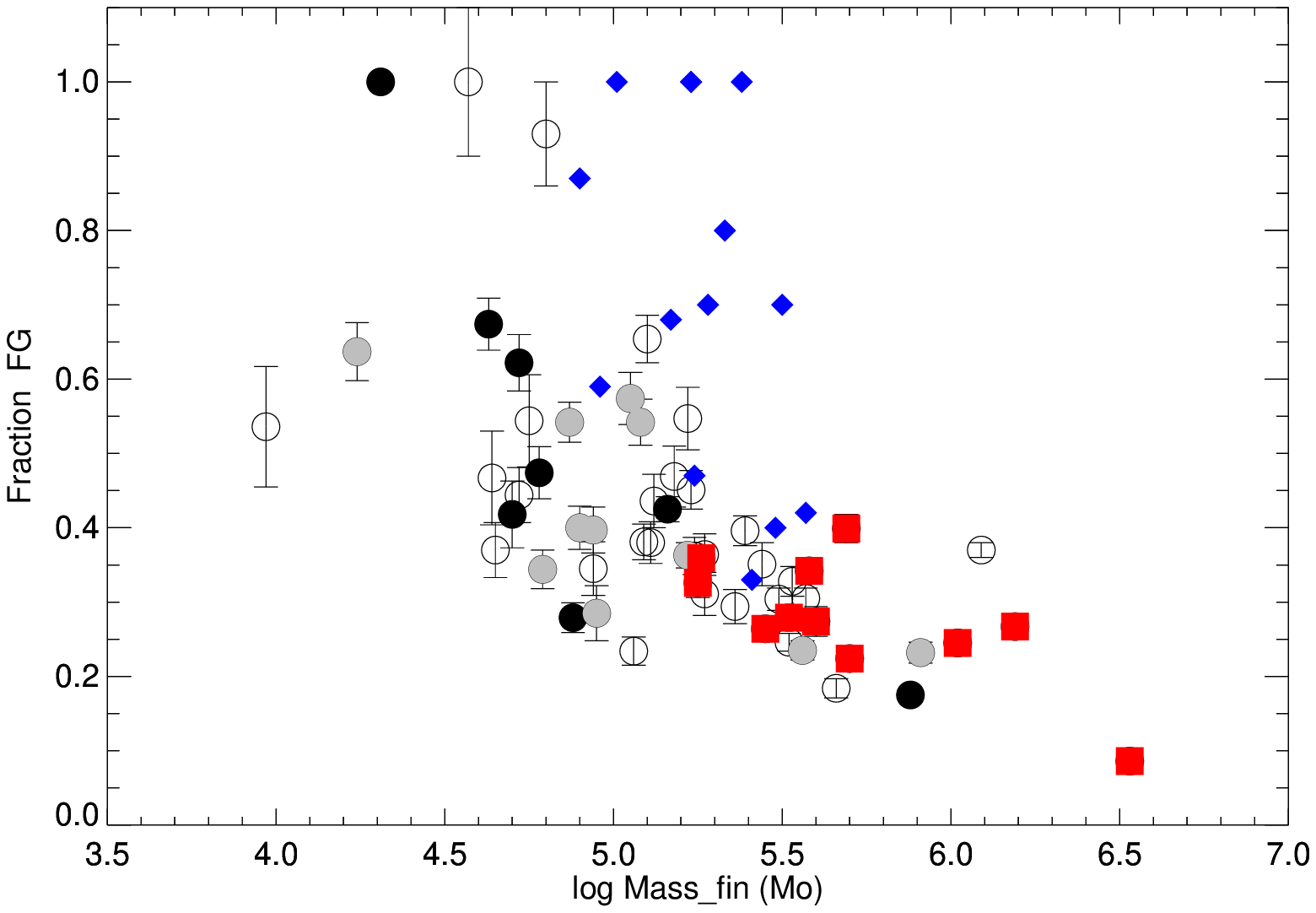}
\includegraphics[width=3.4in]{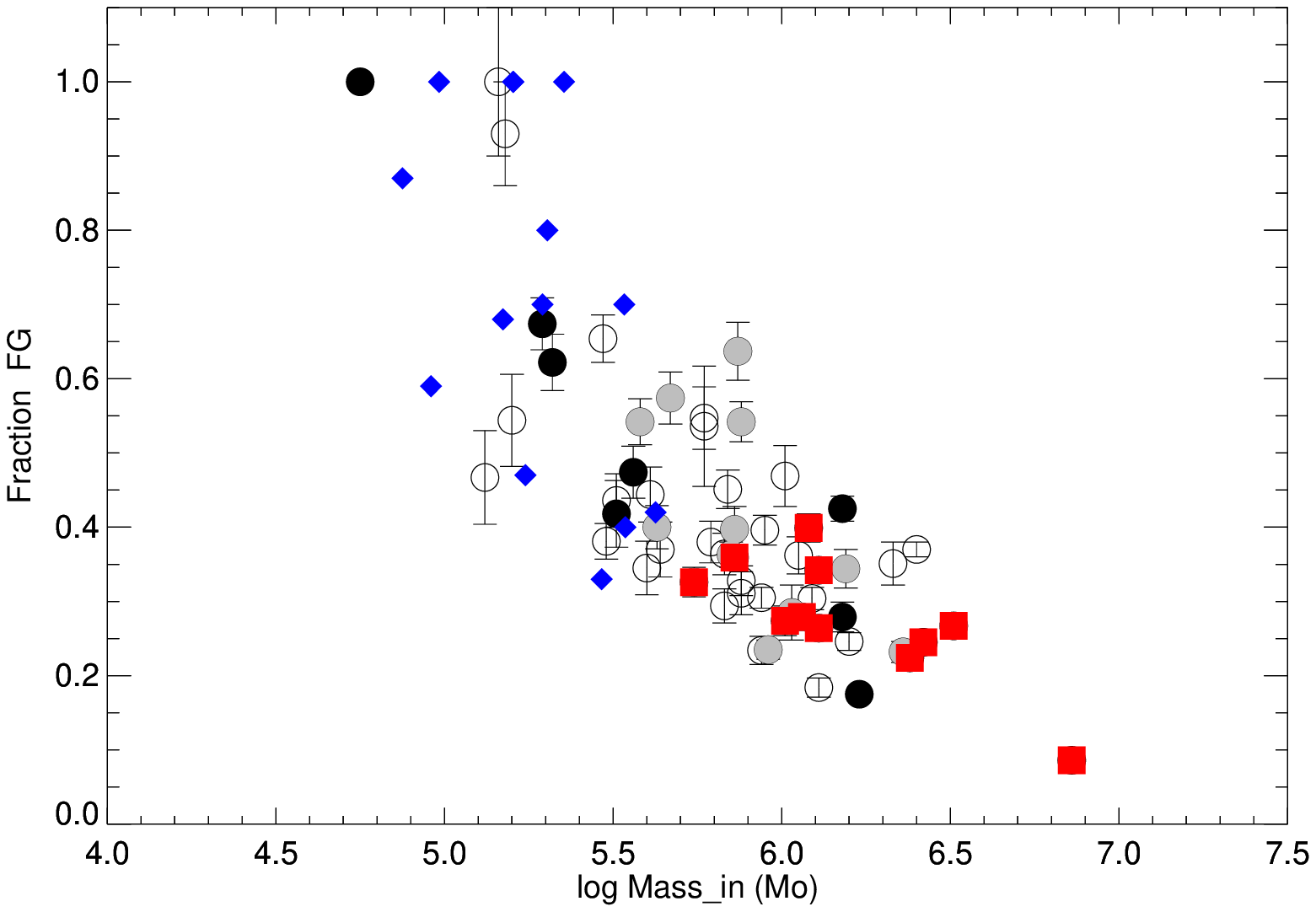}
% \vspace*{-1.0 cm}
 \caption{Upper panel: run of the fraction of FG stars from \citet{Milone2017} with the current (final) cluster mass from \citet{Baumgardt2018}. Symbols are as in Figure~\ref{f:fig1}. The blue filled diamonds are clusters in the Magellanic Clouds (see Gratton et al. 2019 for details). Lower panel: same as upper panel, but with the initial cluster mass from \citet{Baumgardt2019}. See Gratton et al. 2019 for details on how the initial cluster masses are derived for clusters in the Magellanic Clouds. }
   \label{f:fig2}
\end{center}
\end{figure}

\subsection{Relation with globular cluster mass}

\citet{Carretta2010} first noticed that the multiple populations are widespread among globular clusters and that there seems to be a minimum mass for the onset of this phenomenon. This was shown by using different symbols for clusters with and without the Na/O anticorrelation in a diagram where the absolute magnitude of a cluster (a proxy for the globular cluster mass) is plotted against the cluster age. This diagram or a similar version using current mass directly has then been used by many other authors (see e.g. \citealt{Martocchia2018b}). This plot is useful, but does not show two facts that are relevant: (i) the frequency of SG stars is possibly different among different clusters; and (ii) the mass value that should be important for the multiple population phenomenon is not the current one, but rather the mass at the epoch of formation (that we call here initial mass, $M_{\rm in}$). The initial mass is expected to be larger than the current mass, by a factor that depends on the dynamical evolution of the cluster and may be very large.

To overcome these issues, we considered the fraction of first generation stars from  \citet{Milone2017}, supplemented with additional data for Milky Way and Magellanic Cloud clusters, and compared this with the initial and final mass estimates given by \citet{Baumgardt2018} and \citet{Baumgardt2019} (see Figure~\ref{f:fig2}). Even considering the possible limitations in these mass estimates, these plots show that there is a clear correlation between the fraction of SG stars and the cluster mass. This correlation is even more clean when using the initial cluster mass: this was not obvious, because of the difficulties related to the estimate of the initial mass of the clusters. It is important to notice that in this diagram there is no clear distinction between clusters of different metallicity. The simplest interpretation is that while metallicity is important to determine the exact nucleosynthesis involved in the multiple population phenomenon, the basic process that causes the presence of multiple populations is not itself (strongly?) metallicity dependent. This figure also allows to define the minimum mass for the onset of the multiple population phenomenon at about $10^5$~M$_\odot$, though there is considerable scatter in the fraction of FG stars for masses in the range between  $10^5$ and $3\times 10^5$~M$_\odot$, suggesting that in this regime other factors may also be important determining the presence or lack of multiple populations.

\begin{figure}[b]
%\vspace*{-0.6 cm}
\begin{center}
\includegraphics[width=3.0in]{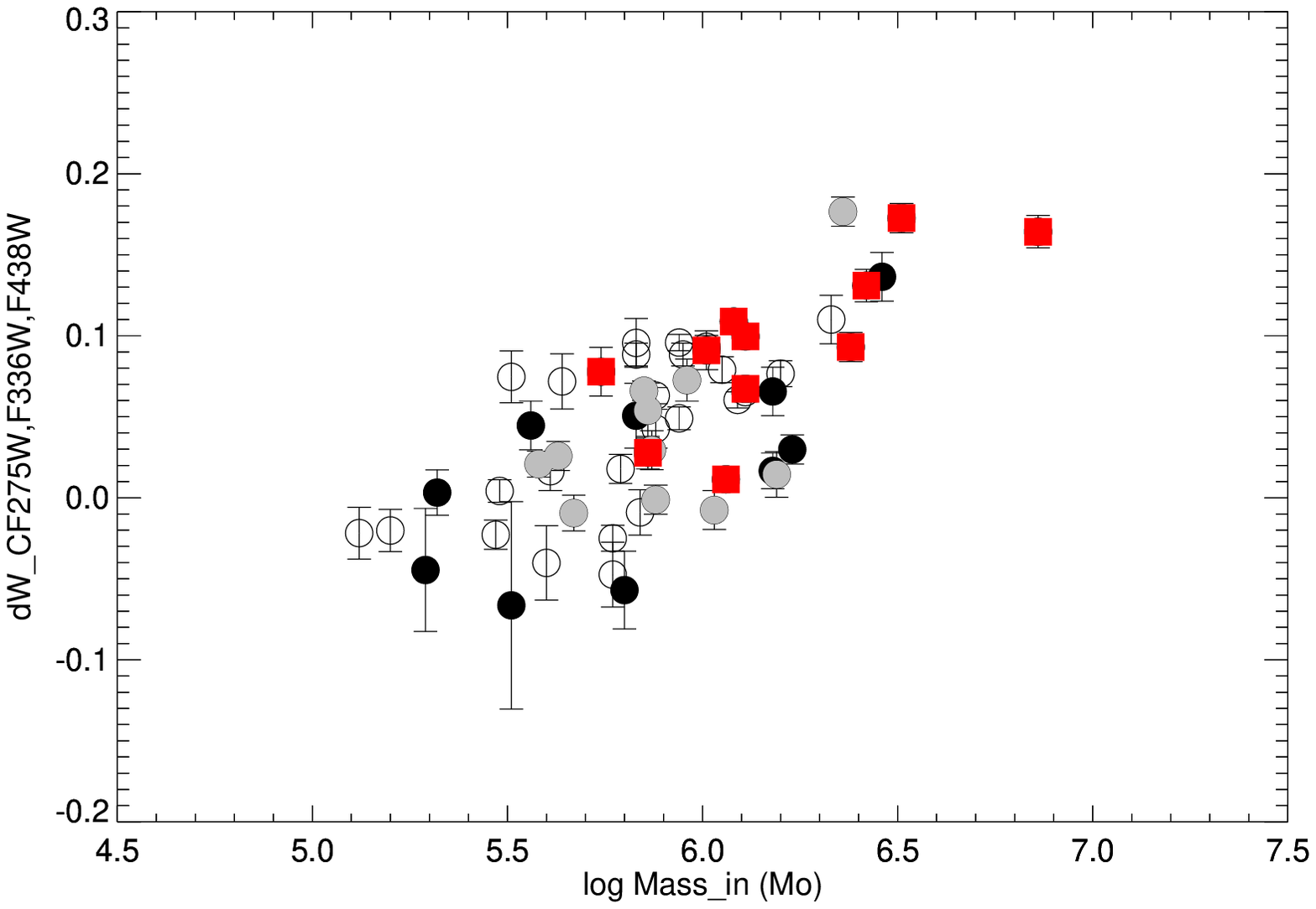}
\includegraphics[width=3.0in]{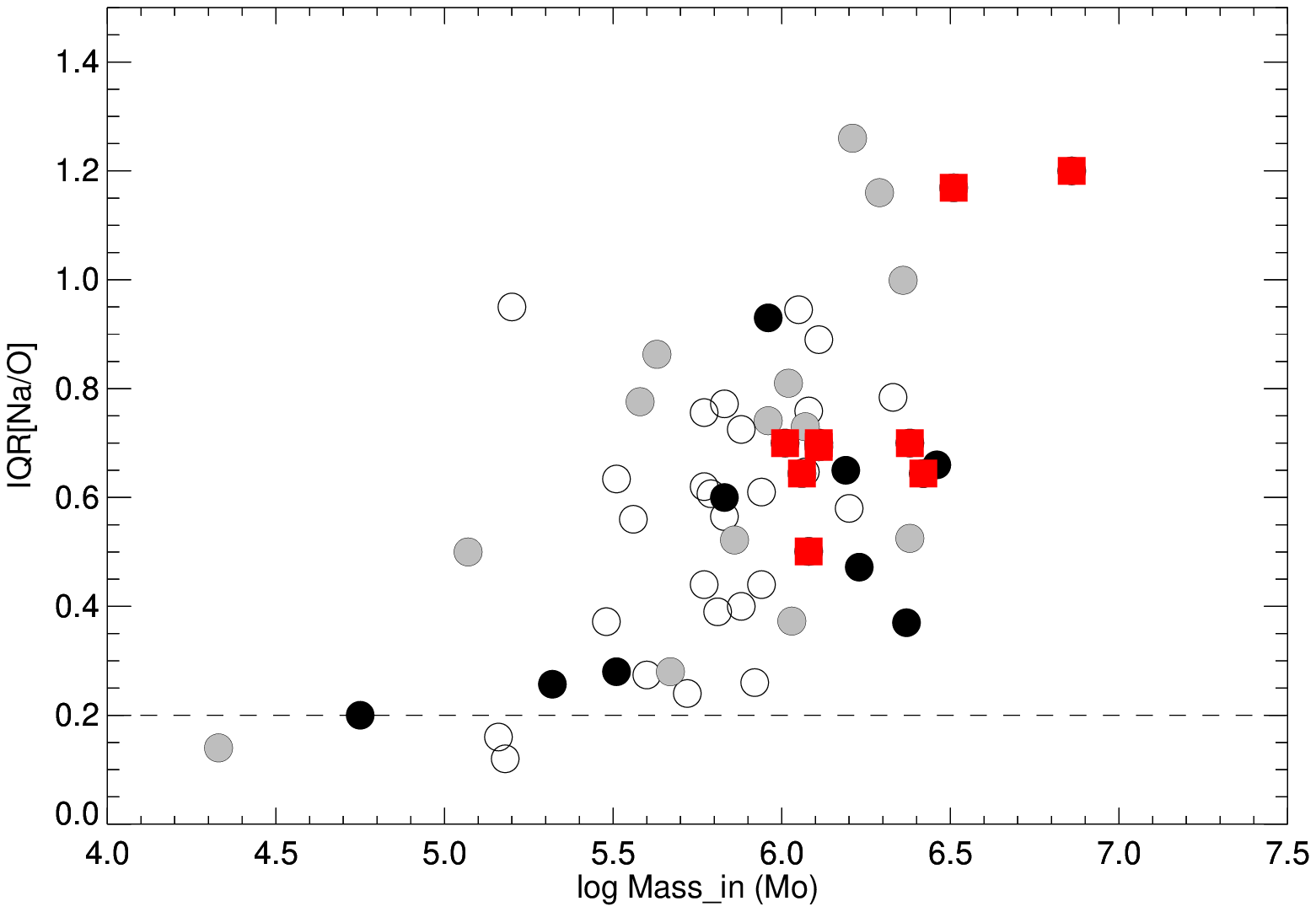}
\includegraphics[width=3.0in]{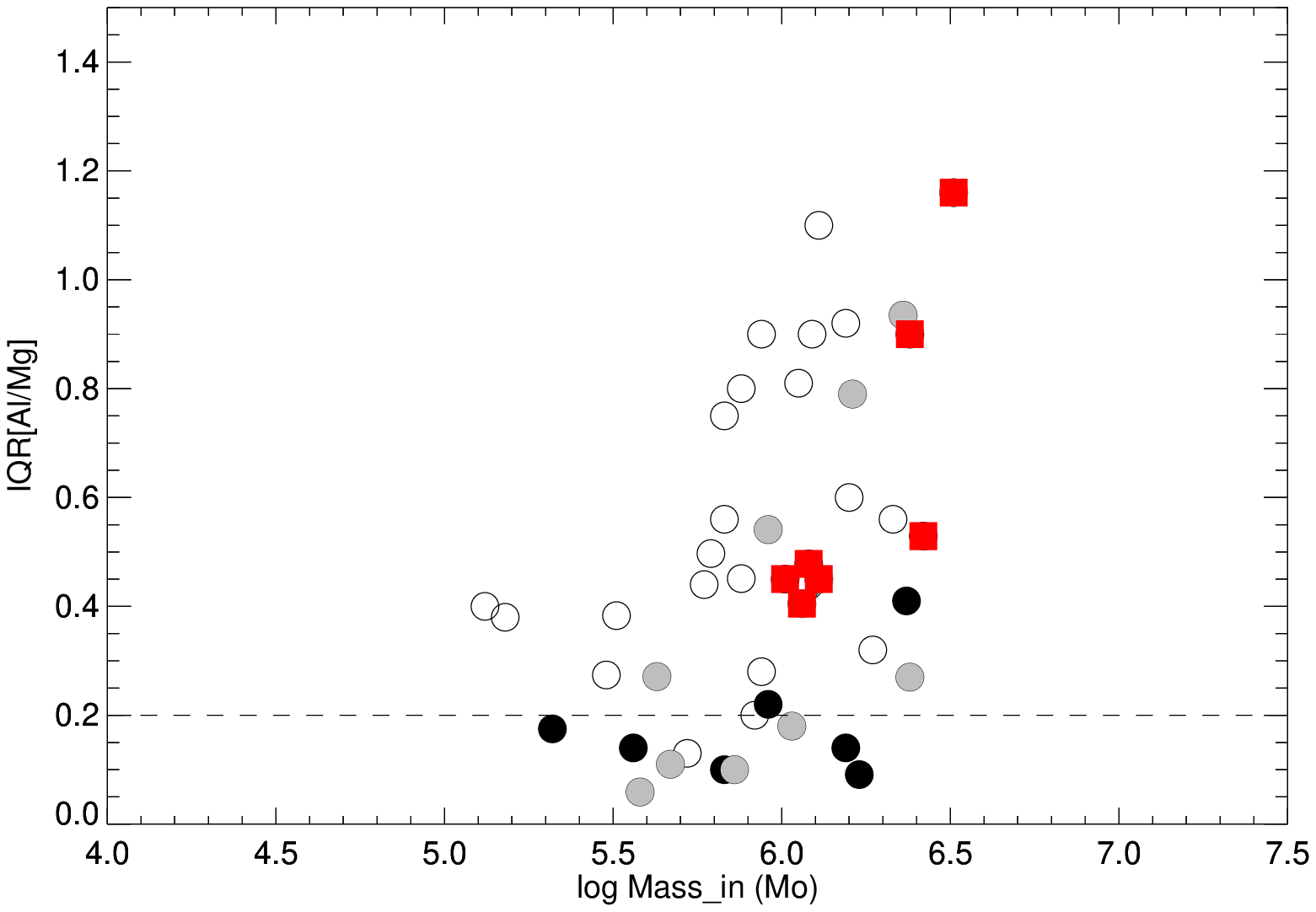}
% \vspace*{-1.0 cm}
 \caption{Upper panel: run of  the photometric index dRGB=$\Delta(275W,F336W,F438W)$ from \citet{Milone2017} - measuring the spread in N abundance within a cluster - with the initial cluster mass from \citet{Baumgardt2018}. Central panel: the same for the IQR[Na/O]. Lower panel: the same for the IGQR[Mg/Al]. Symbols are as in Figure~\ref{f:fig1}. }
   \label{f:fig3}
\end{center}
\end{figure}

Figure~\ref{f:fig3} shows the run of the photometric index dRGB, of IQR[Na/O], and of IQR[Mg/Al] with the initial cluster mass from \citet{Baumgardt2018}. In all cases, there are clear correlations. For the dRGB index (spread in N abundances), there is no strong metallicity dependence and the minimum mass for the onset of the correlation is about $10^5$~M$_\odot$. In the case of the the Na/O and Mg/Al anticorrelation, the minimum masses appear to be larger ($3\times 10^5$~M$_\odot$) and there is a clear dependence on metallicity. The dependence on mass and metallicity were already obtained several years ago (see e.g. \citealt{Carretta2009a, Carretta2009b, Carretta2010, Milone2017}). This is a strong constrain on the mechanism at the origin of the multiple populations because it relates the whole mechanism to the typical burning temperatures, that are likely related to the characteristics of the polluters.

Summarizing, while the fraction of second generation stars (and then the extent of the multiple population phenomenon) does not depend on metallicity, the actual nucleosynthesis is indeed metallicity dependent.

\begin{figure}[ht]
%\vspace*{-0.6 cm}
\begin{center}
\includegraphics[width=3.4in]{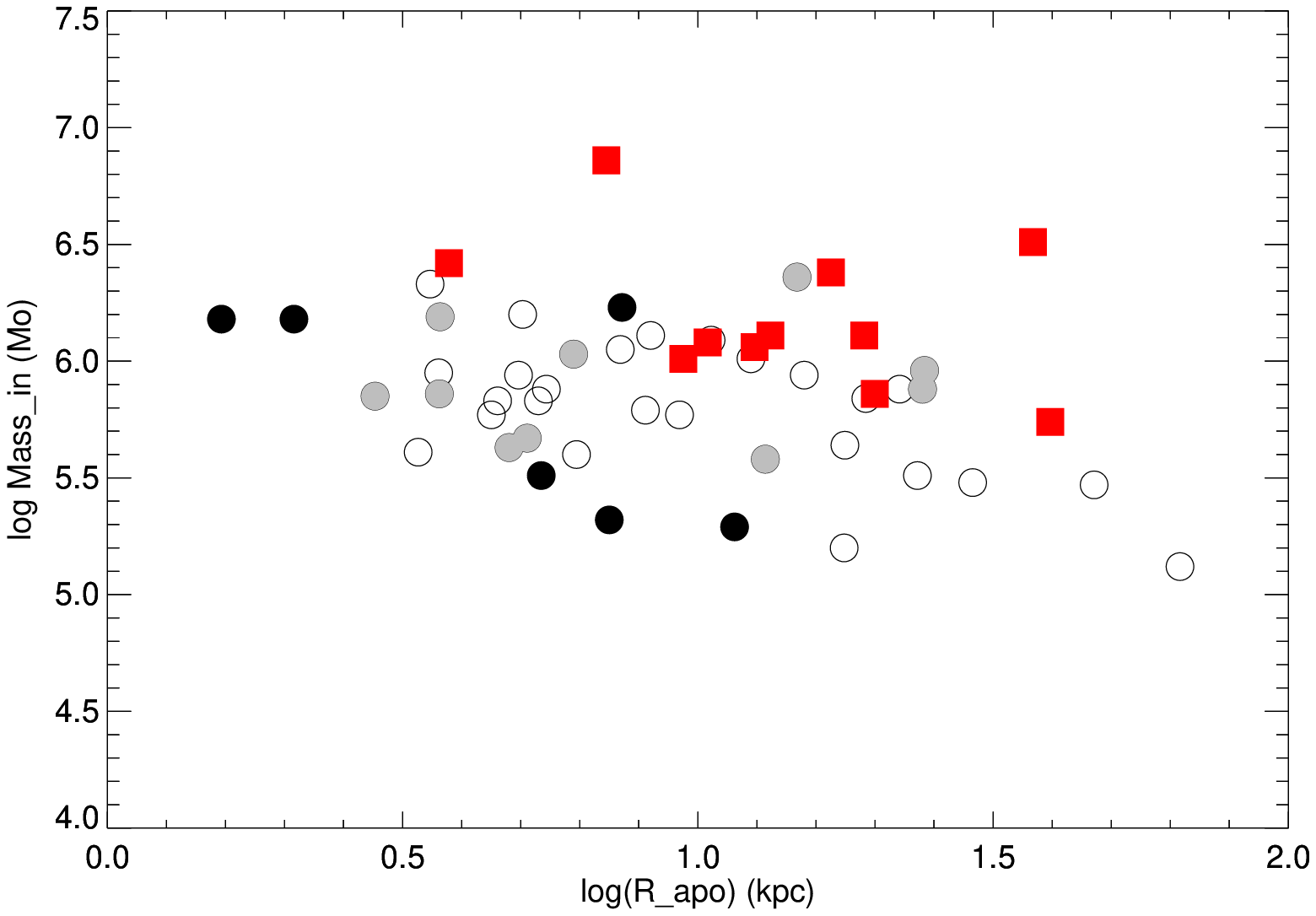}
% \vspace*{-1.0 cm}
 \caption{Run of the initial cluster mass with apo-galactic orbital distance from \citet{Baumgardt2018}. Symbols are as in Figure~\ref{f:fig1}. }
   \label{f:fig4}
\end{center}
\end{figure}

\subsection{Position in the MW}

Another instructive plot is shown in Figure~\ref{f:fig4}, where we display the run of the initial cluster mass with apo-galactic orbital $R_{\rm apo}$\ distance from \citet{Baumgardt2018}. The latter can be considered as a proxy for the distance at which the clusters formed from the center of the Milky Way. We used different symbols for Type~I and II globular clusters. The latter are at the upper envelope of the distribution, that is, they are among the most massive globular clusters at a given $R_{\rm apo}$ and/or they have larger $R_{\rm apo}$ at a given mass, although a bias against smaller clusters may be present as they were not observed by HST and then could not be classified as Type I or II. This supports the view that the more complex internal chemical evolution of type~II globular clusters may be related to their larger mass and/or to the opportunity they had of a longer independent chemical evolution before the interaction with the Milky Way caused the loss of all remaining gas. 

\section{Lithium}

Lithium provides a very sensitive test for both mixing and production/destruction in the polluters. For what concern mixing (at first dredge-up), results show a weak (expected) metallicity dependence in globular cluster stars very similar to field stars \citep{Mucciarelli2012b}. This rules out anomalous mixing and accretion on pre-existing FG stars as explanations for the chemical inhomogeneities related to the multiple populations. 

Various authors (\citealt{Pasquini2005, Lind2009, D'Orazi2010a, D'Orazi2010b, D'Orazi2014, D'Orazi2015, Mucciarelli2011, Monaco2012, gruyters2014}) examined the Li content of FG/SG stars (mainly using subgiant stars). All these studies found very small (if any) difference between Li in FG stars (similar to field) and the bulk of SG stars. There are only a few Li-poor subgiants in globular clusters, generally connected to the extreme second generation stars (the E-population), that is, stars showing extreme Al and presumably He enrichment. There is indeed a good correlation between the fraction of Li-poor stars and that of E-stars among the clusters surveyed so far.

We notice that there is by far too much Li in the majority of SG (Intermediate) stars when compared with the O abundances, even if the ejecta of the polluter are diluted by Li-rich material, a fact already noticed by \citet{Pasquini2005}. This requires production of Li in the polluter, at least for the majority of the SG stars, though not for most of the E-stars. At present, only intermediate mass AGB stars (and novae, that however is not a palatable solution in this context) are known to be able to produce Li. It has been remarked that Li needs to be produced in the polluter at roughly the same level than in the Big Bang nucleosynthesis. It is not clear if this represents a difficulty in the models, because at least some AGB models \citep{D'Antona2012} indeed produce Li at the required level over a quite large range of masses.

Li strongly suggests that massive AGB stars are the best candidate for polluting the material from which the intermediate SG stars formed. However, E-stars – that are the stars with the more extreme pattern of abundances, including very high He - might form from material polluted by other objects. In addition, type I and type II clusters likely have different histories. This calls for a re-examination of the mass budget issue.

\begin{figure}[ht]
%\vspace*{-0.6 cm}
\begin{center}
\includegraphics[width=3.4in]{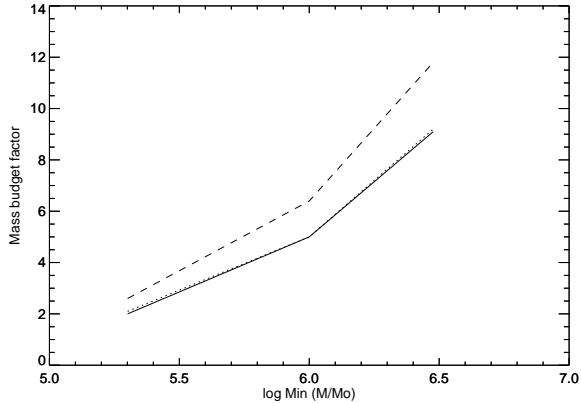}
% \vspace*{-1.0 cm}
 \caption{Estimates of the mass budget factors as a function of the initial cluster mass. Solid, dotted and dashed lines are for initial mass function slopes of 1.7, 2.0, and 2.3, respectively}
   \label{f:fig5}
\end{center}
\end{figure}

\section{Mass budget for clusters of different mass}

In most scenario considered to explain the multiple population phenomenon in globular clusters, only a fraction of the mass of FG stars is used to produce SG stars. Since these are in most cases more numerous than FG stars, the initial mass of the FG (M$_{\rm start}$) should be larger than the mass of the globular cluster at the end of the formation of the SG (the initial mass considered for dynamical evolution, M$_{in}$) to produce enough polluting material. This is known as the mass budget issue. The mass budget factor (M$_{\rm start}$/ M$_{in}$) has been evaluated by various authors according to different scenarios (e.g. \citealt{Carretta2010, Renzini2015, Bastian2018}), with typical values larger than 5. The estimates done so far assume a constant value for the fraction of second generation stars f(SG), and that SG stars are produced by the same mechanism with a universal value for the dilution factor for both intermediate (I) and extreme (E) stars. As discussed above, these assumptions may be incorrect. We then re-estimated the mass budget fraction with values of f(SG) and dilution that are function of the cluster mass. We also assumed that the polluting material incorporated in the I- and E-stars are produced by different mechanisms (see Figure~\ref{f:fig5}). Namely, we assumed here that the polluters for I-stars are intermediate mass AGB stars, and those for the E-stars are fast rotating massive stars (however, this last point is not critical). We considered three different values of the slope of the initial mass function (1.7, 2.0, and 2.3; this last value is the Salpeter one). These values are compatible with current estimates for the slope of the initial mass function in clusters (see \citealt{Beuther2007}; and \citealt{Hosek2019} and references therein). This calculation shows that the mass budget factor needs to be very large ($>5$) only in the most massive globular clusters. Most of these are type II globular clusters, that is, the clusters showing a complex history and likely formed far from the center of the Milky Way.

Finally, we notice that globular clusters retained only a tiny fraction of the ejecta of core collapse supernovae \citep{Suntzeff1996, Renzini2015, Marino2019}. This fraction is $<0.03$\ even for the most massive clusters such as $\omega$~Cen. This is likely due to the depth of the potential well of the proto-cluster nebulae.

\section{Conclusions}

Globular clusters are complex objects that had some internal chemical evolution. They are formed by different generation of stars. Most globular clusters (type I) are mono-metallic and present anti-correlations between the abundances of various elements generated by H-burning at high temperature that follow a repetitive pattern, modulated by mass and metallicity. In most cases they have a moderate mass budget factor. It is possible that a significant fraction of the Type I clusters may be explained by a single polluter class and by a dilution mechanism that is ``built in", such as mass loss from interacting binary stars that have the same mass as the polluters (see e.g. \citealt{Vanbeveren2012}). Massive clusters look more complex: they show a much larger variety and most of them can be classified as type II, that is, they are characterized by many different populations. They also have rather large mass budget factors. The clusters with more complex histories likely formed at larger distances from the center or the Milky Way, and may have formed in satellites of the Milky Way (see e.g. \citealt{Bekki2003}).

Lithium must have been produced in the polluters of the stars with intermediate abundance pattern - that are the vast majority of the second generation stars in the type I clusters. This favours intermediate mass AGB as the polluters. However, the case is different for the stars with more extreme abundance pattern (high Al and He abundances):  in this case the polluting material might come also from other classes of polluters. 

\begin{acknowledgment}
I thank Angela Bragaglia, Eugenio Carretta, Valentina D'Orazi, Sara Lucatello, and Antonio Sollima for revision of the manuscript and important suggestions to its content.
\end{acknowledgment}

%\begin{thebibliography}{}
%\bibliographystyle{unsrtnat}
\bibliographystyle{aa}

\bibliography{Gratton_review.bib}   % name your BibTeX data base

%\end{thebibliography}

\end{document}